\documentclass[twocolumn,twoside]{article}
\usepackage{graphics}

\newcommand{\hb}{\\ \hspace*{2ex}}
\newcommand{\hc}{\\ \hspace*{3ex}}
\begin{document}
\title{DETERMINATION OF DARK MATTER TYPE BY X-RAY SOURCES STATISTICS}
\author{A.V.\,Tugay\\[2mm] 
 Astronomy and Space Physics Department, Faculty of Physics, \\ Taras Shevchenko National University of Kyiv,\hb
 Glushkova ave., 4, Kyiv, 03127, Ukraine,  {\em tugay.anatoliy@gmail.com}
}
\date{}
\maketitle

ABSTRACT.
The current cosmological model includes cold dark matter, which consists of massive nonrelativistic particles. There are also some observational and theoretical evidences for warm dark matter. The existence of warm DM can be examined by measuring of the galaxy clusters density profiles and accurate counting of dwarf galaxies. In this work I suppose that DM haloes are well traced by X-ray gas in clusters, groups, pairs and even single galaxies. The type of DM is inspected with the Xgal sample of 5021 X-ray emitting galaxies observed by XMM-Newton. The selection bias of this sample is also analyzed.

\\[1mm]

{\bf 1. Introduction}\\[1mm]

  Assumptions about the nature of dark matter (DM) can be divided into warm and hot dark matter. Massive neutrino was considered as the main candidate to hot DM up to 2001 and then it was excluded from DM candidates. The difference between cold dark matter (CDM) and warm dark matter (WDM) is concerned with the mass value of a DM particle ($m_{DMP}$). WDM particles became nonrelativistic after nucleosynthesis but before recombination, so their mass should lie in 3 eV - 30 keV interval. In the papers of Boyarsky et al. (2014) and Bulbul et al. (2014) there was found an X-ray emission line that can be the evidence of WDM particle decay with the 7 keV mass. Relativistic DM smooth initial density fluctuations are at any scale less than free-streaming length of a DM particle. Thus, WDM model predicts the lack of dwarf galaxies that can be checked by the astronomical observations. The relation between $m_{DMP}$ and the protogalaxy mass at free-streaming scale ($M_{FS}$) can be derived from Pauli principle (Dodelson, Widrow, 1994): \\[2mm]

\begin{equation}\label{form1}
M_{FS}\equiv \frac{4\pi \rho}{3}\left( \frac{\lambda _\nu }{2} \right) ^3 \approx \frac{3\cdot 10^{15} M_\odot }{\Omega _{DM}} \left( \frac{30 eV}{m_{DMP} } \right) ^2
\end{equation}

We can expect the deficit of galaxies with the mass less than $M_{FS}$. The one from the possibilities to estimate this mass is the consideration of extragalactic X-ray sources statistics. Galaxy clusters, groups and single galaxies can be the sources of the X-ray emission. All these structures are believed to form in a single DM halo. The proportionality of the X-ray luminosity and the mass of a DM halo is assumed in this paper. This allows us to estimate $m_{DMP}$ and the type of DM by the luminosity function of the extragalactic X-ray sources.\\[2mm]

{\bf 2. Sample}\\[1mm]

 The Xgal sample (Tugay, 2013) contains 5021 extragalactic sources from XMM-Newton observations archive. These sources are mostly Seyfert galaxies, galaxy clusters and QSOs (Tugay, 2014); the part of the starburst galaxies is very small (Tugay, Vasylenko, 2011). The sample doesn�t have very large dispersion of the optical luminosities (Fig. 1), so we can find the minimal mass of a DM halo from the constant optical M/L ratio and some limiting value of X-ray luminosity. The following parameters were found for Xgal sample:

\begin{figure}
\resizebox{8.26cm}{!}
{\includegraphics{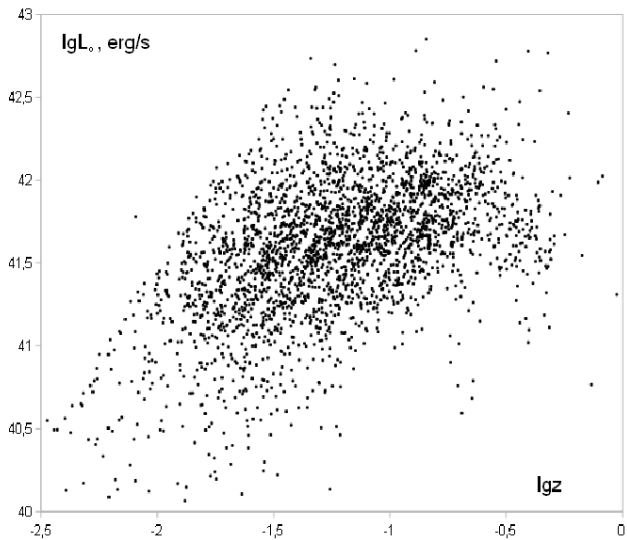}}
\label{hh}
\caption{Distribution of optical u-band luminosities of SDSS galaxies from redshift.}
\end{figure}

1. $r=F_u/F_x$ (see corresponding distributions in Tugay (2012)). The value $lg(r)=2.38\pm 1.07$ was found for the whole sample. The reasons of using this relation instead of the optical fluxes are the selection bias (clearly visible at Fig. 1) and larger absorption in the optical band.

2. The deficit of low luminosity X-ray galaxies can be noticed from Fig. 2. Sources with $F_x<10^{-13}erg/s$ were rejected to avoid the bias from the most luminous ones. Average luminosity $lg(L_X)=40.35\pm 1.25$ was found for the rest of the sample. Limiting luminosity corresponding to $M_{FS}$ and $m_{DMP}$ was estimated as lower bound of 1$\sigma $ interval of $L_X$ distribution. $lg(L_{edge})=39.1\pm 0.6$ was found by this method.\\[2mm]

\begin{figure}
\resizebox{8.26cm}{!}
{\includegraphics{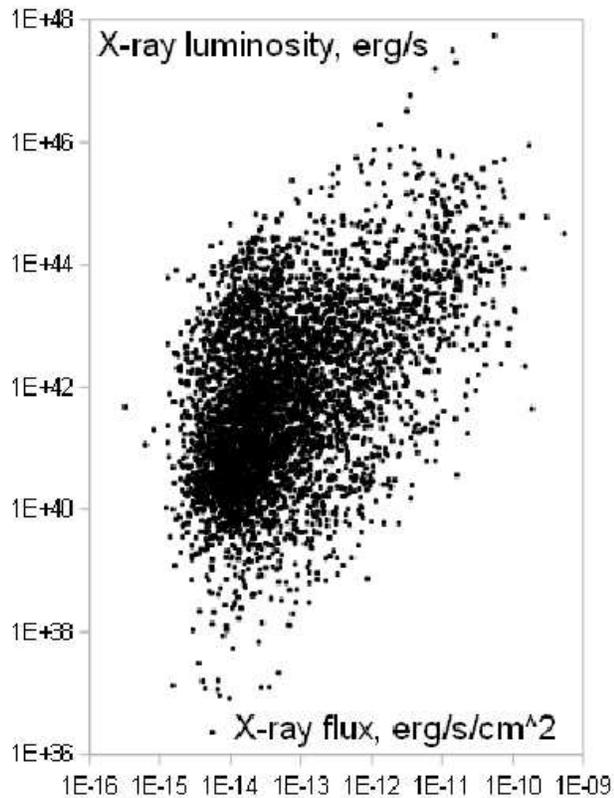}}
\label{hh}
\caption{Distribution of X-ray luminosities of Xgal objects from X-ray flux.}
\end{figure}

{\bf 3. The mass of a WDM particle}\\[1mm]

  The mass of a DM particle was calculated from (1): 

\begin{equation}\label{form2}
m_{DMP}=6 keV \sqrt {\frac{1.2\cdot 10^{45}erg/s}{rKL_{edge}}}
\end{equation}

The optical mass-to-luminosity ratio $K=M/L=140\pm 40$ was taken from (Hradecky et al., 2000) for galaxy clusters. Real ratio may lie in the interval from 70 to 1500 for some clusters, so all estimations are very rough. The resulting value is 

$m_{DMP}=8\div 120 keV;$ $\overline{m_{DMP}}=31 keV$\\[1mm]

{\bf 4. Conclusion}\\[1mm]

Warm and cold dark matter can not be distinguished by the distribution of X-ray luminosities. The main problem seems to be large dispersion of M/L and optical/X-ray ratios. Nevertheless, obtained estimation is close to the results of Boyarsky et al. (2014) and Bulbul et al. (2014) by the order of magnitude.\\[2mm]


{\it Acknowledgements.} The author is thankful to D.A.Iakubovskyi and I.B.Vavilova for the constructive discussion and to O.M.Tugai for spellchecking.
\\[1mm]
{\bf References\\[2mm]}
Boyarsky A. et al: 2014, \hc {\it http://arxiv.org/abs/1402.4119} \\
Bulbul E. et al: 2014, {\it Astrophy. J.,} {\bf 789}, 13. \hc {\it http://arxiv.org/abs/1402.2301} \\
Dodelson S., Widrow L.M.: 1994, {\it Phys. Rev. Lett.,} {\bf 72}, 17. {\it http://arxiv.org/abs/hep-ph/9303287} \\
Hradecky V. et al: 2000, {\it Astrophy. J.,} {\bf 543}, 521. \hc {\it http://arxiv.org/abs/astro-ph/0006397} \\
Tugay A.V., Vasylenko A.A.: 2011, {\it Odessa Astron. Publ.,} {\bf 24}, 72.\\
Tugay A.V.: 2012, {\it Odessa Astron. Publ.,} {\bf 25}, 142. \hc {\it http://arxiv.org/abs/1311.4333} \\
Tugay A.V.: 2013, {\it Adv. Astron. and space Phys.,} {\bf 3}, 116. {\it http://arxiv.org/abs/1311.4337} \\
Tugay A.V.: 2014, {\it IAU Symp.,} {\bf 304}, 168.\\

\vfill

\end{document}